\documentclass[11pt,preprint]{aastex}
\usepackage{epsfig,color}
\usepackage{mathrsfs}
\usepackage{graphicx}
\usepackage{bm}
\usepackage{amsmath}

\def\oiii{[O~{\sc iii}]}
\def\nii{[N~{\sc ii}]}
\def\pp{\prime\prime}
\def\feii{Fe~{\sc ii}}
\def\hei{He~{\sc i}}
\def\heii{He~{\sc ii}}
\def\Ni{[N~{\sc i}]}
\def\fevii{[Fe~{\sc vii}]}
\def\oi{[O~{\sc i}]}
\def\sii{[S~{\sc ii}]}

\shorttitle{Reverberation Mapping of NGC 3516} \shortauthors{Feng et al.}

\begin{document}

\title{Reverberation Mapping of Changing-look Active Galactic Nucleus NGC 3516}

\author{Hai-Cheng Feng\altaffilmark{1,2,3,\bigstar},
Chen Hu\altaffilmark{4},
Sha-Sha Li\altaffilmark{2,4},
H. T. Liu\altaffilmark{1,3,5, \bigstar},
J. M. Bai\altaffilmark{1,3,5, \bigstar},
Li-Feng Xing\altaffilmark{1,3,5},
Wei-Yang Wang\altaffilmark{2,7},
Zi-Xu Yang\altaffilmark{2,4},
Ming Xiao\altaffilmark{4},
Kai-Xing Lu\altaffilmark{1,3,5}}

\altaffiltext{1} {Yunnan Observatories, Chinese Academy of Sciences,
Kunming 650011, Yunnan, People's Republic of China}

\altaffiltext{2} {University of Chinese Academy of Sciences, Beijing
100049, People's Republic of China}

\altaffiltext{3} {Key Laboratory for the Structure and Evolution of
Celestial Objects, Chinese Academy of Sciences, Kunming 650011,
Yunnan, People's Republic of China}

\altaffiltext{4} {Key Laboratory for Particle Astrophysics, Institute of High Energy Physics, Chinese Academy of Sciences, 19B Yuquan Road, Beijing, 100049, People's Republic of China}

\altaffiltext{5} {Center for Astronomical Mega-Science, Chinese Academy of Sciences, 20A Datun Road, Chaoyang District, Beijing, 100012, People's Republic of China}

\altaffiltext{7} {Key Laboratory for Computational Astrophysics, National Astronomical Observatories, Chinese Academy of Sciences, 20A Datun Road, Beijing 100101, People's Republic of China}

\altaffiltext{$^{\bigstar}$}{Corresponding authors: Hai-Cheng Feng, e-mail: hcfeng@ynao.ac.cn,
H. T. Liu, e-mail: htliu@ynao.ac.cn, J. M. Bai, e-mail: baijinming@ynao.ac.cn}

\begin{abstract}
   The changes of broad emission lines should be a crucial issue to understanding the physical properties of changing-look active galactic nucleus (CL-AGN). Here, we present the results of an intensive and homogeneous 6-month long reverberation mapping (RM) monitoring campaign during a low-activity state of the CL-AGN Seyfert galaxy NGC 3516. Photometric and spectroscopic monitoring was carried out during 2018--2019 with the Lijiang 2.4 m telescope. The sampling is 2 days in most nights, and the average sampling is $\sim$3 days. The rest frame time lags of H$\alpha$ and H$\beta$ are $\tau_{\rm{H}\alpha}=7.56^{+4.42}_{-2.10}$ days and $\tau_{\rm{H}\beta}=7.50^{+2.05}_{-0.77}$ days, respectively. From a RMS H$\beta$ line dispersion of $\sigma_{\rm{line}} = 1713.3 \pm 46.7 \/\ \rm{km \/\ s^{-1}}$ and a virial factor of $f_{\sigma}$ = 5.5, the central black hole mass of NGC 3516 is estimated to be $M_{\rm{BH}}= 2.4^{+0.7}_{-0.3} \times 10^{7} M_{\odot}$, which is in agreement with previous estimates. The velocity-resolved delays show that the time lags increase towards negative velocity for both H$\alpha$ and H$\beta$. The velocity-resolved RM of H$\alpha$ is done for the first time. These RM results are consistent with other observations before the spectral type change, indicating a basically constant BLR structure during the changing-look process. The CL model of changes of accretion rate seems to be favored by long-term H$\beta$ variability and RM observations of NGC 3516.

\end{abstract}

\keywords{Active galaxies (17); Photometry (1234); Reverberation mapping (2019); Seyfert galaxies (1447); Spectroscopy (1558); Supermassive black holes (1663)}

\section{Introduction}
Seyfert galaxies, a subset of active galactic nuclei (AGNs), were first reported by \citet{Se43}. Depending on the
relative widths of Balmer and forbidden lines, \citet{KW74} separated the Seyfert galaxies into two subclasses, type 1
and type 2. The Balmer lines in type 1 are broader than the forbidden lines, while all the lines in type 2 have
comparable widths. Additionally, some sources with strong narrow lines and weak broad lines are classified
as types 1.8 and 1.9 \citep{Os81}. In the framework of unification scheme \citep{An93}, the various types of AGNs are
caused by the different torus orientation with respect to the line of sight (LOS). The observational evidence of this concept is that some spectropolarimetric observations detected the hidden broad emission lines (BELs) in some type 2
AGNs \citep[e.g.,][]{AM85, MG90, Tr92}. However, the deeper spectropolarimetry found numerous type 2 AGNs without broad components, and the absence of spectropolarimetric BELs challenged our understanding of the exact physical processes \citep[e.g.,][]{Tr03, Wu11}. A handful of AGNs, so-called changing-look AGNs (CL-AGNs), may shed light on some intrinsic properties. The term of "changing-look" originated from X-ray observations in which sources changed from Compton-thin
to Compton-thick on timescales of years, or vice versa \citep[e.g.,][]{Ma03,La15,La17}. The optical spectroscopic observations indicate that these objects are usually accompanied by a significant change in spectral
type, i.e., the BELs disappear or reappear, and a number of subsequent CL-AGNs are identified by optical spectrum \citep[e.g.,][]{St93, Ar99, EH01, De14, Ki18, Wa18, Sh19}.

The mechanisms of CL-AGNs are still debated. Early explanations are mainly the change in obscuration \citep{Bi05},
while some recent studies favor the change of accretion rate \citep{St18a,Sn20}. Photoionization researches
show that the CL behavior in CL quasars can be fully explained by the photoionization responses of the BELs to the
extreme variability of the ionizing continuum \citep{Gu20}. Besides, several objects can be interpreted as the
tidal disruption events \citep[TDEs,][]{Me15,Ri20}. All the models can generate significant variance of the observed intensity of continuum and emission lines. These models might be restricted by the statistical studies after
extending the sample of CL-AGNs. Several groups have devoted to searching for new CL-AGNs, and tens candidates
have been discovered based on the large optical and X-ray surveys, such as the Sloan Digital Sky Survey
\citep[SDSS; e.g.,][]{La15,Ma16,Ru18,Ho20}, intermediate Palomar Transient Factory \citep[iPTF;][]{Ge17}, Pan-STARRS1 \citep[PS1;][]{Ma19}, Catalina Real-time Transient Survey \citep[CRTS;][]{Ya18}, and XMM-Newton slew survey \citep{Ze18,Ko20}. In the future, the Large Synoptic Survey Telescope (LSST) will regularly monitor millions of AGNs,
and the number of CL-AGNs will further increase. The evolution of the AGN type can be well studied if we obtain
continuous multi-wavelength data of the complete changing process, and the development of time-domain surveys makes it possible to achieve this. The brightening of NGC 2617 \citep{Sh14} and 1ES 1927+654 \citep{Tr19} triggered the alert
of the All-Sky Automated Survey for Supernovae (ASAS-SN\footnote{http://www.astronomy.ohio-state.edu/asassn}), and
the follow-up observations confirmed these two CL-AGNs. However, so far there are only small available datasets for
most detected CL-AGNs, especially around the critical regime of transition between types 1 and 2. Moreover, previous
studies rarely gave the change of geometry and kinematics in broad-line region (BLR).

To date, the GRAVITY interferometer can directly resolve the spatial profile of BLR, but only 3C 273 has been observed \citep{St18b}. Another powerful technique to investigate the size, geometry, and kinematics of the BLR is reverberation mapping \citep[RM, e.g.,][]{BM82,Pe93}. The technique is based on a widely accepted assumption that the BELs are generated through photoionization driven by the central continuum radiation. Thus, a delayed correlation and a time lag
of $\tau$ (in the rest frame) will be expected between the light curves (LCs) of AGN continua and BELs. The time lag is caused by the light-travel time, and the size of BLR is $R_{\rm{BLR}} = c \tau$, where $c$ is the light speed. Based on
the assumption of virialized motion of gas clouds in BLR, the virial mass of the central black hole is expressed as
\begin{equation}
  M_{\rm{BH}} = f \frac{c \tau v^{2}}{G},
\end{equation}
where $G$ is the gravitational constant, $v$ is the BEL velocity width, and $f$ is the virial factor depending on the geometry and kinematics of the BLR \citep{Pe93}. $f$ can be determined with the $M_{\bullet}-\sigma_{\ast}$ relation for
the low-$z$ inactive galaxies, quiescent galaxies, and/or reverberation mapped AGNs, where $\sigma_{\ast}$ is the stellar velocity dispersion of galaxy bulge \citep[e.g.,][]{Tr02,On04,Pi15,Wo13,Wo15}. Recently, \citet{Li17} proposed a new method to measure $f$ using the gravitationally redshifted broad emission lines. Several known CL-AGNs have been studied with RM, including Mrk 590 \citep{Pe98}, NGC 3516 \citep{De10}, NGC 2617 \citep{Fa17}, and NGC 4151 \citep{De18}. \citet{Ok19} confirmed NGC 1566 to be a CL Seyfert galaxy. Since the BELs are weak during type 2, all the previous RM campaigns were
only carried out in type 1 state. Therefore, the difference of BLR before and after the change in type is still unclear.

NGC 3516 (z = 0.00884) is one of the first reported Seyfert galaxy characterized by the complex emission lines. The Balmer lines show a clear double-peak component which is usually regarded as the contribution of the outer parts of an accretion disk \citep[e.g.,][]{Po02, St17}, but the detailed structure of emitting region is still under debate. In the first five months of 1990, \citet{Wa93} carried out a spectroscopic monitoring of the object with 21 spectra, and gave a time lag of $\tau_{\rm{H}\beta}$ = $\rm{7^{+3}_{-3}}$ days. A high cadence RM campaign in 2007 gave $\tau_{\rm{H}\beta}$ = $\rm{7.43^{+1.99}_{-0.99}}$ --- $11.68^{+1.02}_{-1.53}$ days \citep{De10}. During the first half of 2012, \citet{De18} obtained a time lag of $\tau_{\rm{H}\beta}$ = $\rm{8.11^{+0.75}_{-0.58}}$ days, which is well consistent with the previous results. Unlike other AGNs, e.g., NGC 5548 with variable BLR size \citep[e.g.,][]{Pe02,Lu16,Ko18}, the H$\beta$ time lag of NGC 3516 is basically constant over 20 yrs. The long-term monitoring campaign confirmed that the object is a CL-AGN after 2014 \citep{Sh19}. The BELs nearly disappeared, and the continuum declined by more than a factor of 2. The X-ray observations of $Suzaku$ also detected a faint phase in the period of 2013--2014 \citep{No16}. Interestingly, \citet{Il20} reported a new optical flare and several coronal lines at the end of 2019, indicating that the object may switch on again. The repeat spectral transition in such short timescales hints that its CL phenomenon is probably driven by changes in accretion rate. \citet{Kr02} found eight absorption troughs in the UV band, and this supports the outflows in NGC 3516. \citet{Du18b} found a clear change in one of the absorption troughs using the $HST$/COS spectrum in 2011, and it was interpreted as bulk motion.

The data used in this paper are from a new RM campaign which aims to investigate the disk-like BLRs. We present the general results of RM and discuss the origin of changing-look. The structure of BLR will be studied in the next paper. The information of observations and data reduction are described in Section 2; the LCs, time-series analysis, black hole mass, and velocity-resolved delays are presented in Section 3; Discussion is in Section 4; and Summary is in Section 5.

\section{Observations and Data Reduction}
From November 2018 to May 2019, the observations of NGC 3516 were carried out with Yunnan Faint Object Spectrograph and Camera (YFOSC) mounted at the 2.4 m optical telescope. The telescope is located at Lijiang Observatory, Yunnan Observatories, Chinese Academy of Science. The detailed information of telescope, instruments, and Observatory were described in \citet{Wa19} and \citet{Xi20}. The observing procedure is similar to that in \citet{Fe20}. Here, we briefly describe the main points. The cadence is 2 days in most nights, and two small gaps in the beginning of the observations are caused by the bad weather. In each photometric night, we would take a broadband Johnson $B$ image before or after the spectroscopic exposure, and 59 data points were successfully observed during our monitoring campaign. Four nearby comparison stars are always in the field of view (FOV) which can provide accurate calibration of the target ($\leq$1\%). The 69 spectra were obtained by Grism 14, which provides a dispersion of 1.76 \AA\ pixel$^{-1}$ and a wavelength coverage of 3600--7460 \AA. Considering the seeing condition ($\sim$1$^{\pp}_{\cdot}$5) of the telescope, we adopted a long slit with a projected width of 2$^{\pp}_{\cdot}$5, which can minimize the effects of atmospheric differential refraction and provide a relatively high spectral resolution. For each spectroscopic observation, we simultaneously put the target and a comparison star in the slit to improve the flux calibration. We also take a spectrophotometric standard star in the night to calibrate the absolute flux of the comparison star. Besides, we employed a UV-blocking filter to eliminate the secondary spectrum. Therefore, the data around H$\alpha$ can be used in our analysis. To evaluate the effects of the seeing and grism, we also took two additional spectra using a wide slit width (5$^{\pp}_{\cdot}$05) and low resolution Grism 3 (2.93 \AA\ pixel$^{-1}$) in different nights. We found that the two spectra of comparison star are consistent with each other, and they are also consistent with the spectra obtained by Grism 14 in good weather conditions. Therefore, we averaged the two spectra as a fiducial spectrum which can be used to calibrate the flux of target.

All the raw photometric and spectroscopic data were reduced with the standard IRAF software. The aperture radius of photometry was 6$^{\pp}$, and then the differential magnitudes were obtained using the photometric comparison stars.
The one-dimensional spectra of target and comparison star were extracted from a uniform aperture radius of 5$^{\pp}_{\cdot}$943. \citet{vW92} calibrated the spectral flux using the \oiii\ emission line, and this choice is
widely adopted in many RM campaigns \citep[e.g.,][]{Be09b,Hu16,Fa17,Du18a}. Generally, this method needs wide slit widths
for extended sources. The \oiii\ emitting region of NGC 3516 is larger than 10$^{\pp}$ \citep[see also Figure 2 in][]{Sh19}. Thus, the \oiii-based calibration approach is not suitable for our observations. All the spectra were calibrated by the spectrum of comparison star. We corrected the Galactic extinction of the spectra using the dust map of \citet{Sc98} and the extinction curve of \citet{Fi99}. Finally, the mean spectrum is shown in the upper panel of Figure 1. We also measured the rms spectrum (see the bottom panel of Figure 1) using the same method as in \citet{Pe04}. Note that the variance of seeing and calibration accuracy of wavelength will affect the shapes of narrow lines \citep[see also][]{Pe04}, we therefore subtracted the fitted narrow lines and host components before our calculation of the RMS spectrum (see details in Section 3.3).

\section{Results and Analysis}
\subsection{Light Curves}
The $B$-band photometric LC is shown in the upper panel of Figure 2, and the error bars are calculated as in \citet{Li19}, i.e., the combination of Poisson errors (from the target and comparison stars) and systematic errors (from
the variation of comparison stars). The photometric data are listed in Table 1. The continuum (rest-frame 5100 \AA) LC is measured from the median value between 5090 \AA\ and 5110 \AA. We found that the standard deviations ($\sigma_{\rm{s}}$)
in the continuum window are consistent with the corresponding Poisson errors. Thus, we adopted $\sigma_{\rm{s}}$ as the errors of continuum. For the emission-line fluxes, we first inspected the mean and rms spectra to confirm the continuum windows around each line (6400--6415 \AA\ and 6760--6780 \AA\ for H$\alpha$, 4780--4800 \AA\ and 5090--5110\AA\ for H$\beta$). We then subtracted the continuum through a linear fit. The fluxes of H$\alpha$ and H$\beta$ (both broad and narrow components) were obtained by integrating 6460--6650 \AA\ and 4810--4910 \AA, respectively. The Poisson errors
of each line can be directly measured by the IRAF. However, both continuum and emission lines should be affected by the systematic error which might be related to weather condition and instrument state. Following the method in \citet{Du14},
we smoothed the LCs with a median filter of 5 data points, and then subtracted the original LCs. The systematic errors are estimated through the standard deviation of the residuals. The systematic errors are usually much larger than the Poisson errors due to the variation of host galaxy flux in the slit. The LCs of H$\alpha$, H$\beta$, and continuum are shown in Figure 2, and the corresponding data are listed in Table 1.

\subsection{Time-series Analysis}
As shown in Figure 1, the starlight of host galaxy heavily contaminates the brightness of 5100 \AA, and the narrow slit width will introduce some unexpected uncertainties because of the variance of seeing condition. We hence use the photometric ($B$-band) data as the continuum. To estimate the light-travel time lag between continuum and lines, we employed the interpolated cross-correlation function \citep[ICCF,][]{WP94}, one widely accepted method. We found that the location
of ICCF peak ($r_{\rm{max}}$) is consistent with the location of ICCF centroid above 0.8$r_{\rm{max}}$, and only the centroid value was adopted in our time lag analysis. Another standard method, Z-transformed discrete correlation function \citep[ZDCF,][]{Al97}, was also used to measure the time lags of H$\alpha$ and H$\beta$. The ZDCF gives results similar to those derived from the ICCF. The results of the ICCF in our analysis are used to estimate $M_{\rm{BH}}$. The time dilation is corrected via $\tau_{\rm{rest}} = \tau_{\rm{obs}}/(1 + z)$, where $z=0.00884$ \citep{Ke96}.

In order to estimate the uncertainties of time lag, we employed the ``flux randomization/random subset selection" (FR/RSS) method \citep{Pe04} to build up the cross-correlation centroid distribution (CCCD). Each CCCD is generated by
10000 Monte Carlo realizations. The lower and upper errors of lag were determined by the 15.87\% and 84.13\% quantiles of the CCCD, respectively. The ICCFs and CCCDs are shown in the right panels of Figure 2. The time lags of H$\alpha$ and H$\beta$ are measured using their integrated LCs, and $\tau_{\rm{rest}} \approx 7.5$ days (see Table 2). Our results are consistent with the previous RM results of H$\beta$ \citep{Wa93,De10,De18}. For H$\alpha$ and H$\beta$, \citet{Sh19} got
the time lags of $\sim$ 15--17 days, which are larger than our RM results. Also, the H$\alpha$ time lag of $14 \pm 2$ days obtained in \citet{Wa93} is larger than our result of 7.5 days. However, the average sampling of the RM observations in \citet{Wa93} is $\sim$ 7 days, and the average sampling in \citet{Sh19} is $\sim$ 3 months for H$\alpha$ and $\sim$ 2 months for H$\beta$. Our average sampling is $\sim$3 days, and then our measured time lags of 7.5 days seem more reliable.

\subsection{Contamination of Host Galaxy}
The contribution of host galaxy might affect the RM results due to the extremely faint phase of NGC 3516 during our observations \citep{Sh19}. Therefore, it is important to evaluate the influence of the host galaxy on LCs. \citet{Fe17} found that the variance of seeing will change the brightness in the aperture for the extended source. We examined the correlation between brightness and seeing of photometry, and did not find any significant correlation. Thus, the host
galaxy should marginally affect our $B$-band photometric LC because of the large aperture radius (6$^{\pp}$). In terms of the spectroscopy, we fitted each spectrum using Levenberg-Marquardt least-squares minimization technique which is widely used to decompose the starlight \citep{Ba15,Hu16}. The detailed fitting procedure is described below.

\begin{enumerate}
  \item The fitting model consists of a single power-law continuum ($f_{\lambda}$ = $C \lambda^{\alpha}$), broad optical \feii\ template from \citet{Do08}, host galaxy template of \citet{BC03}, one Gaussian for broad \hei\ $\lambda$5876 emission, two double-Gaussians for broad H$\alpha$ and H$\beta$, tens single-Gaussians for narrow permitted lines and forbidden lines (including several coronal lines). The permitted narrow lines include H$\alpha$, H$\beta$, \heii\ $\lambda$4686, and \hei\ $\lambda$5876, while the forbidden lines are \oiii\ $\lambda\lambda$4959,5007 doublet, \Ni\ $\lambda$5199, \fevii\ $\lambda$5721, \fevii\ $\lambda$6086, \oi\ $\lambda\lambda$6300,6363 doublet, \nii\ $\lambda\lambda$6548,6583, and \sii\ $\lambda\lambda$6718,6732. Throughout our fitting process, all the narrow components are assumed to have the same profile, and the fitting range covers 4400--6800 \AA\ in the rest-frame.

  \item First, we fit the mean spectrum to obtain an optimal host template. The flux ratios of \nii\ doublet $\lambda$6583/$\lambda$6548 and \oiii\ doublet $\lambda$5007/$\lambda$4959 are fixed to the theoretical values of 2.96 and 3, respectively. The template of host galaxy is fixed (except resolution) in the following fitting of individual spectrum, and we also fix the relative flux ratios of other narrow lines to the mean spectrum. Although, \citet{Il20} found the variability in the coronal lines, they should not affect our fitting results. If we fix the spectral index (i.e., $\alpha$) to the best-fitting value of the mean spectrum, the scatter in the LCs can be moderately reduced. The spectra of $HST$ indicate that different Balmer lines (H$\alpha$, H$\beta$, and H$\gamma$) of NGC 3516 have the nearly same profile \citep{De16}. In the fitting, the broad H$\alpha$ and H$\beta$ are assumed to have the same redshift and width, and this will generate a better fitting result.

  \item We also fit the spectra using three to four Gaussians, and the host galaxy template is replaced with the spectrum of host component obtained by the principal component analysis (PCA) method \citep{Il20}. The final results from different ways are consistent with each other.

\end{enumerate}

As an example, the fitted mean spectrum is depicted in Figure 3. Figure 4 shows the LCs of continuum and emission lines measured from the best-fitting model. The AGN continuum is measured from the fitted power law around 5100 \AA, and the fluxes of H$\alpha$ and H$\beta$ are obtained by the broad components. Both AGN and comparison star are point sources,
while host galaxy is an extended source. Thus, the effects of seeing condition are different for AGN and host galaxy.
The flux of AGN can be well calibrated by the comparison star, and the host galaxy will contribute some fake variability
in continuum and narrow lines. Comparing Figure 2 and Figure 4, the fitted results of H$\alpha$ and H$\beta$ are almost identical with the integrated results, but there is smaller data point scatter in the fitted continuum LC. The continuum
are directly contaminated by the strong host galaxy, and this means that even small fluctuations in the host galaxy will generate large scatter in its LC. The fluxes of integrated H$\alpha$ and H$\beta$ are mainly affected by the narrow emission lines. However, the proportion of narrow lines to broad lines is much smaller than the ratio of host galaxy to AGN continuum, resulting in a small scatter in the BEL LC. Although, the spectral fitting can basically remove the contribution of host galaxy, it can also introduce extra errors. This might generate a worse LC in the case where there is weak effect
of host galaxy \citep[e.g.,][]{Hu20}. Indeed, the LC of fitted H$\beta$ (only contaminated by narrow H$\beta$) shows slightly larger scatter, but the LCs of fitted H$\alpha$ and continuum show smaller scatter. We found that the lags calculated from the fitted LCs are the same as the integrated results. This indicates that the contribution of host galaxy might marginally influence our RM analysis.

\subsection{Black Hole Mass}
$M_{\rm{BH}}$ can be estimated via Equation (1). Two widely used methods are employed to measure the velocity width of BEL, including the full width at half-maximum (FWHM) and the line dispersion ($\sigma_{\rm{line}}$) of BEL. Here, only $\sigma_{\rm{line}}$ is adopted to measure $M_{\rm{BH}}$ due to the irregular BELs. The finite resolution of the instrument is corrected using the FWHM of \oiii. Comparing the \oiii\ FWHM ($\sim$747 km s$^{-1}$) in our mean spectrum and intrinsic width of 250 km s$^{-1}$ \citep{Wh92}, we can obtain a spectral resolution of 704 km s$^{-1}$. Following \citet{Pe04}, we calculated the width $\sigma_{\rm{line}}$ of each broad line from both RMS and mean spectra. The results of $\sigma_{\rm{line}}$ are given in Table 2. \citet{On04} gave $f_{\sigma}$ = 5.5, which is in agreement with the value of \citet{Wo10}. Thus, we derived four values of $M_{\rm{BH}}$ for NGC 3615 (see Table 2). Our measured black hole masses are consistent with the previous RM measurements \citep{Pe04,De10,De18,Sh19}.

\subsection{Velocity-resolved Lags}
The lags calculated from the LCs of total flux of BELs only represents the mean scale of BLR. However, the detailed geometry and kinematics of BLR are still ambiguous. We can separately measure the lags of different velocity regions (the so-called velocity-resolved RM) to recover the information of BLR \citep[e.g.,][]{Ho04}. The two-dimensional velocity-delay map
will be provided in our future work. Here, we simply divided the line flux into ten bins in velocity space. Each bin contains equal flux in the RMS spectrum. The LCs of each bin are shown in Figure 5, and the corresponding lags are plotted in Figure 6. Both H$\alpha$ and H$\beta$ show clear shorter lags towards the red side, and this trend is consistent with that obtained by H$\beta$ in \citet{De10}. The time lag of the blue side of double-peaked H$\alpha$ is larger than that of the red side, by $\Delta \tau_{\rm{H\alpha}}\approx 20$ days (see Figure 6). For double-peaked H$\beta$, $\Delta \tau_{\rm{H\beta}}\approx 10$ days.

\section{Discussion}
From the comparison of the H$\beta$ profile observed in 1943 by \citet{Se43} and in 1967 by \citet{An68}, it can
be clearly seen that the broad H$\beta$ component was present and absent in the epochs of 1943 and 1967, respectively \citep{Sh19}. From 1990 to 2007, NGC 3516 showed dramatic BELs, and three RM campaigns had been carried out during this period \citep{Wa93,De10,Sh19}. \citet{De18} provided a RM data of H$\beta$, obtained during 2012, for NGC 3516. Both mean and rms spectra show a significant broad H$\beta$ component \citep[see Figure 4 in][]{De18}, which means that NGC 3516 may be a type 1 AGN in 2012. From 2014 to 2018, NGC 3516 was in states with very weak H$\beta$ BELs \citep{Sh19}. Reliable lags of BELs are detected by the high-quality data from our RM campaign of NGC 3516 from November 2018 to May 2019. Thus, NGC 3516 may be the first CL-AGN which has the RM data before and after changing its type. These H$\beta$ variability and RM results may provide some clues to the BLR, and the mechanism of the changes in type.

According to \citet{La15}, the obscuration timescale can be estimated as the crossing time for an intervening dust cloud orbiting outside a BLR
\begin{equation}
t_{\rm{cross}}= 0.07 \left( \frac{r_{\rm{orb}}}{1 \rm{lt-day}} \right)^{3/2} M_{8}^{-1/2} \arcsin \left( \frac{r_{\rm{src}}}{r_{\rm{orb}}} \right) \/\ \rm{yr},
\end{equation}
where $r_{\rm{orb}}$ is the orbital radius of the foreground cloud on a circular, Keplerian orbit around the central black hole, $M_{8}$ is the black hole mass in units of $10^8 M_{\odot}$, and $r_{\rm{src}}$ is the true siz of the BLR. The foreground cloud should be placed at $r_{\rm{orb}}\geqslant 3 r_{\rm{BLR}}$, so that it can cover a substantial portion
of the gas in the BLR, specifically gas at low velocities that contributes flux to the cores of the line profile \citep{La15}. Seyfert 1 galaxy PGC 50427 has a ratio of $r_{\rm{dt}}/r_{\rm{BLR}} = 2.4$ \citep{Po15}, where $r_{\rm{dt}}$ is a dust torus radius. The rest-frame 5100 \AA\ AGN luminosities at the epochs of the H$\beta$ BLR lag measurement and
the dust lag measurement are $L(\rm{5100 \AA})\approx 10^{43} \/\ \rm{erg \/\ s^{-1} \/\ cm^{-2}}$ for NGC 5548 with $r_{\rm{dt}}/r_{\rm{BLR}} =4.4$ \citep{KM20}. \citet{Ma20} found that $r_{\rm{dt}}$ is about a factor of 5.3 times larger than $r_{\rm{BLR}}$ for Seyfert 1 galaxy Z229-15. The dust torus innermost radii are generally larger compared to the H$\beta$ BLR radii, by a factor of $\sim$ 4 for dust reverberation-mapped Seyfert galaxies \citep{KM20}. NGC 3516 has $r_{\rm{dt}}=52.5$ lt-days with $L(\rm{5100 \AA})=10^{42.63} \/\ \rm{erg \/\ s^{-1} \/\ cm^{-2}}$ \citep{KM20}, and $r_{\rm{BLR}}=7.5$ lt-days with $L(\rm{5100 \AA})=10^{42.75} \/\ \rm{erg \/\ s^{-1} \/\ cm^{-2}}$ obtained in our observations. Thus, it is appropriate to using $r_{\rm{dt}}=52.5$ lt-days and $r_{\rm{BLR}}=7.5$ lt-days in Equation (2). Combining $M_{8}=0.24$ for the H$\beta$ BEL, $t_{\rm{cross}}=3.9$ yr for NGC 3516, i.e., the ``disappearance" timescale
of the H$\beta$ BEL is about 4 yr.

The orbital period of the dust cloud $P_{\rm{orb}}=0.228 M_{8}^{-1/2} \left( r_{\rm{dt}}/1 \rm{lt-day} \right)^{3/2}$
yr $=177$ yr for NGC 3516. The ratio of appearance to ``disappearance" of the H$\beta$ BEL is $(P_{\rm{orb}}-t_{\rm{cross}})/t_{\rm{cross}}=44$. NGC 3516 has the H$\beta$ BEL from 1997 to 2007 \citep[see Figure 12 in][]{Sh19}, and during the first half of 2012 \citep{De18}. The ``disappearance" of the H$\beta$ BEL is from 2014 to 
2018 \citep[see Figure 12 in][]{Sh19}. Our observations of NGC 3516 from November 2018 show the H$\beta$ BEL, as well 
as the H$\alpha$ BEL, again. Assuming its appearance from 1997 to 2012, and its ``disappearance" from 2013 to 2018, 
the ratio of appearance to ``disappearance" of the H$\beta$ BEL is 3, which is much smaller than 44. If assuming its appearance from 1990 to 2012, and then the ratio is 4.4, which is also much smaller than 44. If NGC 3516 was always a Seyfert 1 galaxy after 1967 and before 2013, i.e., the upper time limit of the H$\beta$ BEL appearance is from 1968 to 
2012, and then the ratio is $\lesssim$ 8.8, which is smaller than 44. The above results are only true for a 
single cloud. If considering ten clouds with similar orbital elements that might obscure the BLR on the same time scale, $(P_{\rm{orb}}-10t_{\rm{cross}})/10t_{\rm{cross}}=3.5$, which is not inconsistent with the roughly estimated ratios of  3--8.8. In addition, $t_{\rm{cross}}\approx 4$ yr is roughly consistent with the H$\beta$ BEL ``disappearance" timescale 
of about 5 yr. These results significantly depend on the baseline and sampling of spectroscopic observations. The available spectral data could not determine whether the obscuration model of CL-AGNs might be favored or disfavored for NGC 3516. The appearance and ``disappearance" timescales of the H$\beta$ BEL, and their ratio may be used to test the obscuration model of CL-AGNs by the future spectroscopic monitoring with a sampling of several months.

\citet{Es97} proposed a two-zone disk model, including an advection-dominated accretion flow within a transition radius $R_{\rm{tr}}$ and an outer standard thin disk. A change of $R_{\rm{tr}}$ may generate a transition in the accretion regime. \citet{Al20} find the similarity in the X-ray spectral evolution between CL-AGNs and black hole X-ray binaries, which implies that the observed CL-AGN phenomena may be related to the state transition in accretion physics. The hard X-ray emission from NGC 3516 \citep[e.g.,][]{Il20} requires an additional component than a standard thin disk, such as corona, advection-dominated accretion flow or hot accretion flow \citep[see review by][]{YN14}. X-ray observations of NGC 3516 show photon indices of $\Gamma =1.75^{+0.01}_{-0.02}$ on 2013 May 23--24, and $\Gamma =1.72^{+0.08}_{-0.12}$ on 2009 October 28 \citep{No16}. On average, the photon flux on 2013 May 23--24 is higher by a factor of about 2 than that on 2009 October 28 \citep[see Figure 2 in][]{No16}. This flux difference and the consistent $\Gamma$ indicate that the obscuration model of CL-AGNs might not be favored for NGC 3516. The latest research shows that X-ray variability observed in CL-AGNs
is produced by accretion disk structure and accretion rate changes \citep{Ig20}. The latest photoionization calculations produce a natural sequence of the successive weakening of H$\alpha$ and H$\beta$, when the ionizing continuum decreases \citep{Gu20}. As the BELs in NGC 3516 are strong, or NGC 3516 is a Seyfert 1 galaxy, the optical spectrum shows the appearance of a strong excess of continuum shortward of 4200 \AA\, compared to that in ``disappearance" of the BELs \citep{De16,Sh19}. This strong excess is contributed by the pure-AGN with the strong BELs, because the host galaxy spectrum will be very low in the same wavelength regime \citep[see Figure 3 in][]{Sh19}. Considering the X-ray results, the appearance of this strong excess naturally implies a high accretion rate of the central black hole, e.g., a standard thin disk, and consequently the relevant BELs are stronger. The main difference between the obscuration origin and the accretion rate origin of CL-AGNs is likely the dust torus radius changing with the AGN ultraviolet-optical luminosity. The luminosity decrease of CL-AGN Seyfert 1 galaxy Mrk 590 is interpreted as an intrinsic change in the mass accretion rate of black hole \citep{De14}. The innermost radius of the dust torus in Mrk 590 decreased rapidly after the AGN ultraviolet-optical luminosity dropped, and $r_{\rm{dt}}(\rm{faint})/r_{\rm{dt}}(\rm{bright})\sim 0.3$  between the faint and bright phases \citep{KM20}. Thus, the changes of accretion rate of black hole, switching between different accretion modes, might be favored as the CL origin in NGC 3516.

The strong host galaxy might weaken the intrinsic relative variation amplitudes of the AGN continuum. When the AGN is weaker than the host galaxy, this weakening will obviously influence the CCFs between the 5100 \AA\ and BEL LCs, and then the relevant time lags of BELs. Figure 3 shows a strong host galaxy contribution ($>$ 60\%) in the continuum of NGC 3516. The relative variation amplitudes of the 5100 \AA\ integrated LCs are smaller than those of the 5100 \AA\ best-fit LCs (see Figures 2 and 4). The $B$ band photometric LCs reflect the 5100 \AA\ variations of the pure AGN, because the seeing effect on the photometric LCs is slight due to the large photometric aperture that basically reaches the border of the host galaxy. However, the seeing effect on the spectroscopic 5100 \AA\ LCs is serious because of the narrow slit and the brightness distribution of the host galaxy. Ultimately, the host galaxy influences the spectroscopic 5100 \AA\ LCs as the AGN is weaker than the host galaxy. The influence of the host galaxy can be weakened in the spectral decomposition best-fit LCs, but can not be completely eliminated. Thus, the $B$ band LCs have more clear structures than the 5100 \AA\ LCs in Figures 2 and 4, and the CCF between the $B$ band and 5100 \AA\ LCs in Figure 4 is better than that in Figure 2. These clear structures are important to get the reliable time lags of BELs through the CCF analysis. That was the reason why the 5100 \AA\ continuum LCs are replaced with the photometric LCs in the RM researches of weaker AGNs with the stronger host galaxy.

In our RM campaign of NGC 3516, H$\beta$ shows a lag of $\tau_{\rm{H}\beta} = \rm{7.50^{+2.05}_{-0.77}}$ days, which is
well consistent with the previous RM results (see also Section 1). The measured radii of BLR in NGC 3516 are basically constant for about 30 yrs. According to the photoionization model of BELs and the radius--luminosity relationship of AGNs established by the RM observations and \citep[e.g.,][]{Ka00,Be09a,Be13}, the BLR sizes of H$\alpha$ and H$\beta$ should change with the continuum luminosity variations. Our measured values of $r_{\rm{BLR}}$ for NGC 3516 are basically consistent with those derived in \citet{De10}. However, the 5100 \AA\ and H$\beta$ fluxes in our observations are lower than those obtained in 2007 \citep{De10}. It seems that $r_{\rm{BLR}}$ did not change with the luminosity variations. Our RMS
spectrum gives a H$\beta$ velocity dispersion of 1713.3 $\pm$ 46.7 km s$^{-1}$, which is consistent with $\sigma_{\rm{rms}}(\rm{H}\beta)= 1591 \pm 10$ km s$^{-1}$ obtained in \citet{De10}. Also, it seems that line width $v$ did not change with the luminosity variations. Moreover, our values of $r_{\rm{BLR}}$ are consistent with the results
of the other two RM observations in 1990 \citep{Wa93} and 2012 \citep{De18}. Furthermore, our RM observations show $\tau_{\rm{H}\alpha}$/$\tau_{\rm{H}\beta} = 1$ for NGC 3516. If BLR was a narrow belt, it would be easy to understand
these above results. Based on the photoionization assumption of BELs, $r_{\rm{BLR}}$ of a narrow belt should not change as the 5100 \AA\ continuum (as a proxy of ionizing continuum) varies, because there is not enough space to change the position of the optimal photoionization zone when the continuum intensity varies. In terms of BLR, NGC 3516 might be very similar to NGC 1097, in which the elliptical ring is very narrow \citep{Er95}. However, no RM observation was run during the dim state from 2014 to 2018 \citep[e.g.,][]{Sh19}, and then we cannot determine whether $r_{\rm{BLR}}$ is equal for the bright and
the dim states. Thus, the suggestion of a narrow belt of BLR needs to be tested by the RM observations during period of being Seyfert 2 galaxy with the larger telescopes.

The velocity-resolved RM of H$\alpha$ is done for the first time in our RM observations of NGC 3516. Our results show
that the lag of H$\alpha$ is the same as that of H$\beta$, and the velocity-resolved lags are also consistent with each other (see Figure 6), indicating the same geometry and kinematics of BLR for H$\alpha$ and H$\beta$. Figure 6 shows that
the lags gradually increase from 3000 km s$^{-1}$ to -3000 km s$^{-1}$, which is generally regarded as the signature of inflow. Also, this velocity-resolved lag trend might be generated by an elliptical disk-like BLR, or a circular disk-like BLR plus a spiral arm-like BLR. The double-peaked H$\beta$ and H$\alpha$ BELs appear in the mean spectra (see Figure 3)
and the rms spectra (see Figure 6), and have $\tau_{\rm{H}\alpha}$/$\tau_{\rm{H}\beta} = 1$. Double-peaked BELs were predicted and fitted well by the elliptical disk model \citep{Er95}. Thus, the BLR in NGC 3516 might be an elliptical disk-like one. The geometry and kinematics of the BLR during our observation period might be consistent with
those obtained in \citet{De10} due to the similar velocity-resolved H$\beta$ lag maps measured in our and their works.
Under pure recombination, all the broad Balmer lines are predicted to originate from the same region, but in practice,
the Balmer lines of most AGNs usually show a stratified structure which may be caused by the optical depth effects within
the BLR, i.e., the optical depth of H$\alpha$ is largest, followed by H$\beta$, and so on through the Balmer series \citep[e.g.,][]{Ne75,Re89,Ko04,Be10}. \citet{Be10} found a mean ratio of $\tau_{\rm{H}\alpha}/ \tau_{\rm{H}\beta} = 1.54$ for 13 nearby Seyfert 1 galaxies, and $\tau_{\rm{H}\alpha}>\tau_{\rm{H}\beta}$ except for Mrk 142 (see their Table 13). Also, the consistent lags of the Balmer lines were detected in some other AGNs, such as NGC 4051 and Mrk 374 \citep{Fa17}, PKS 1510-089 \citep{Ra20}. Whether the lags of the Balmer lines are consistent with each other might constrain the physical conditions within BLR of CL-AGN, e.g., the BLR structure. If the ratio of lags is equal to one, it supports the
idea of a narrow ring for the BLR because there isn't enough space for the differences in lags to occur that are expected from photoionization models. The radius of the narrow ring BLR we proposed will influence the radius--luminosity relation \citep{Ka00,Be13}.

If the major axis of the ellipse is in the plane of the sky (perpendicular to the observer's line of sight, i.e., the major axis orientation angle $\varphi_{0}=90^{\circ}$ or $270^{\circ}$ \cite[see Figure 2 in][]{Er95}), the maximum line-of-sight velocity $v_{\rm{los}}(\rm{max})$ of an elliptical ring BLR, corresponding to the red and blue wings of the BEL, will have a time delay that matches the radial distance from the black hole. The mean spectra in Figure 3 and the rms spectra in Figure 6 show that the red wing could reach higher velocities than the blue wing (relative to their narrow line components or the cores of the lines) for the H$\beta$ and H$\alpha$ BELs. Thus, the redward $v_{\rm{los}}(\rm{max})$ is larger than the blueward $v_{\rm{los}}(\rm{max})$ for the H$\beta$ and H$\alpha$ BELs, and the time lag of the blue wing is larger than that of the red wing, which indicate $\varphi_{0}=270^{\circ}$. The clouds around the pericenter and apocentre in an elliptical ring BLR will produce the redward and blueward $v_{\rm{los}}(\rm{max})$, respectively, as $\varphi_{0}=270^{\circ}$. The above results are qualitatively consistent with the velocity-resolved time lag trends presented in Figure 6. If $\varphi_{0}$ deviates heavily from $\varphi_{0}\sim 270^{\circ}$, the observed lag trends might be generated by an elliptical ring BLR, but, $v_{\rm{los}}(\rm{max})$ would not match the radial distance from the black hole. Different combinations of parameters, BLR inclination angle $i$, $\varphi_{0}$, eccentricity $e$, pericenter distance, and etc, could produce different profiles of double-peaked BELs \citep{Er95}. The inflow also could produce the observed lag trends in Figure 6. For an axisymmetric disk-like inflow BLR, the time lag difference $\Delta \tau$ between the red and blue wings will give $ 2r_{\rm{BLR}}\sin(i) \approx c \Delta \tau /(1+z)$. Also, the time lag $\tau_{\rm{zero}}$ of the line core might give $r_{\rm{BLR}} \approx c \tau_{\rm{zero}}/(1+z)$, and there might be $c(\tau_{\rm{zero}}-\tau_{\rm{red}})/(1+z) = r_{\rm{BLR}}\sin(i)$, i.e., $\sin(i) = (\tau_{\rm{zero}}-\tau_{\rm{red}})/\tau_{\rm{zero}}$, for the time lags of the red wing and the line core. Thus, there might be $i \sim 15^{\circ}$ if an axisymmetric disk-like inflow BLR is in NGC 3516.

\section{Summary}
In this paper, we present photometric ($B$-band) and spectroscopic monitoring results of CL-AGN NGC 3516 using the Lijiang 2.4 m telescope. This object should be the first CL-AGN which has complete RM data before and after the change in type. From our analysis, we conclude the following results.
\begin{enumerate}
\item Our RM observations give a homogeneous and good sampling data set of NGC 3516 during a low-activity phase. Our sampling of RM data in most nights is 2 days, and the average sampling is $\sim$3 days, which are smaller than the estimated time lags of H$\alpha$ and H$\beta$.

\item The H$\alpha$ lag is $\tau_{\rm{H}\alpha}$ = $\rm{7.56^{+4.42}_{-2.10}}$ days, and the H$\beta$ lag is $\tau_{\rm{H}\beta}$ = $\rm{7.50^{+2.05}_{-0.77}}$ days, which are consistent with each other. The Balmer lines may originate from the same region of BLR. A black hole mass of $M_{\rm{BH}}= 2.4^{+0.7}_{-0.3} \times 10^{7} M_{\odot}$ is obtained using a RMS H$\beta$ line dispersion of $\sigma_{\rm{line}}=1713.3 \pm 46.7 \/\ \rm{km \/\ s^{-1}}$ and $f_{\sigma}$ = 5.5.

\item The time lags increase towards negative velocity for both H$\alpha$ and H$\beta$. The velocity-resolved RM of H$\alpha$ is done for the first time. The line width and velocity-resolved reverberation signature of H$\beta$ in our observations are consistent with the RM results of H$\beta$ in \citet{De10}.

\item The CL model based on changes of accretion rate of black hole seems more favored for NGC 3516.

\end{enumerate}

NGC 3516 might be the first CL-AGN which has the RM data before and after changing its type. The mystery of geometry and kinematics of the BLR in NGC 3516 needs the further observational researches of BELs. A new RM campaign will be run during the next new observation period with the Lijiang 2.4 m telescope.

\acknowledgements{We are very grateful to the anonymous referee, as well as Dragana Ilic and Luka Popovic, for constructive comments leading to significant improvement of this paper. We also thank the helpful discussions of Yu-bin Li. JMB thanks the financial support of the National Natural Science Foundation of China (NSFC; grant No. 11991051). We thank the joint fund of Astronomy of the NSFC and the Chinese Academy of Sciences (CAS), under grants No. U1831125 and U1331118. KXL acknowledges the financial support from the NSFC (grants No. 11703077 and 12073068) and the Yunnan Province Foundation (202001AT070069). HTL thanks the financial support of the CAS Interdisciplinary Innovation Team. We acknowledge the support of the staff of the Lijiang 2.4 m telescope. Funding for the telescope has been provided by Chinese Academy of Sciences and the People's Government of Yunnan Province.}

\clearpage

\begin{deluxetable}{ccccc}
  \tablecolumns{5}
  \setlength{\tabcolsep}{5pt}
  \tablewidth{0pc}
  \tablecaption{LCs of Photometry and Lines}
  \tabletypesize{\scriptsize}
  \tablehead{
  \colhead{JD - 2458000}             &
  \colhead{Mag}                      &
  \colhead{JD - 2458000}             &
  \colhead{H$\alpha$}                &
  \colhead{H$\beta$}
} \startdata
431.423843 & 2.226 $\pm$ 0.004 & 447.351852 & 106.09 $\pm$ 6.16 & 9.95 $\pm$ 0.85 \\
433.405000 & 2.235 $\pm$ 0.004 & 448.412002 & 105.74 $\pm$ 6.16 & 11.40 $\pm$ 0.84 \\
451.415313 & 2.238 $\pm$ 0.005 & 451.419306 & 99.15 $\pm$ 6.16 & 10.35 $\pm$ 0.83 \\
452.427616 & 2.235 $\pm$ 0.005 & 466.446250 & 112.74 $\pm$ 6.16 & 10.90 $\pm$ 0.84 \\
466.441007 & 2.282 $\pm$ 0.006 & 474.455787 & 94.85 $\pm$ 6.16 & 9.27 $\pm$ 0.84 \\
... & ... & ... & ... & ... \\
\enddata
\tablecomments{\footnotesize The fluxes of H$\alpha$ and H$\beta$ are in units
of 10$^{-14}$ erg s$^{-1}$ cm$^{-2}$. (This table is available in its entirety
in machine-readable form.)}
\label{Table1}
\end{deluxetable}

\begin{deluxetable}{ccccccccccc}
  \tablecolumns{6}
  \setlength{\tabcolsep}{5pt}
  \tablewidth{0pc}
  \tablecaption{Rest-frame Lags, Line Widths, and BH Masses}
  \tabletypesize{\scriptsize}
  \tablehead{
  \colhead{Line}                &
  \colhead{}                    &
  \colhead{Lag (days)}          &
  \colhead{}                    &
  \multicolumn{3}{c}{$\sigma_{\rm{line}}$ (Km s$^{-1}$)}  &
  \colhead{}                    &
  \multicolumn{3}{c}{$M_{\rm{BH}}$ ($\times 10^{7}M_{\odot}$)} \\ \cline{5-7} \cline{9-11}
  \colhead{}                    &
  \colhead{}                    &
  \colhead{}                    &
  \colhead{}                    &
  \colhead{Mean}                &
  \colhead{}                    &
  \colhead{RMS}                 &
  \colhead{}                    &
  \colhead{Mean}                &
  \colhead{}                    &
  \colhead{RMS}
} \startdata
H$\alpha$ & & $\rm{7.56^{+4.42}_{-2.10}}$ & & 2256.8 $\pm$ 16.8 & & 1823.8 $\pm$ 41.3 & & $\rm{4.1^{+2.4}_{-1.1}}$ &  & $\rm{2.7^{+1.6}_{-0.8}}$   \\
H$\beta$ & & $\rm{7.50^{+2.05}_{-0.77}}$ & & 1881.5 $\pm$ 13.8 & & 1713.3 $\pm$ 46.7 & & $\rm{2.8^{+0.8}_{-0.3}}$ &  & $\rm{2.4^{+0.7}_{-0.3}}$   \\
\enddata
\label{Table2}
\end{deluxetable}

\clearpage

\begin{figure*}
 \begin{center}
  \includegraphics[scale = 0.8]{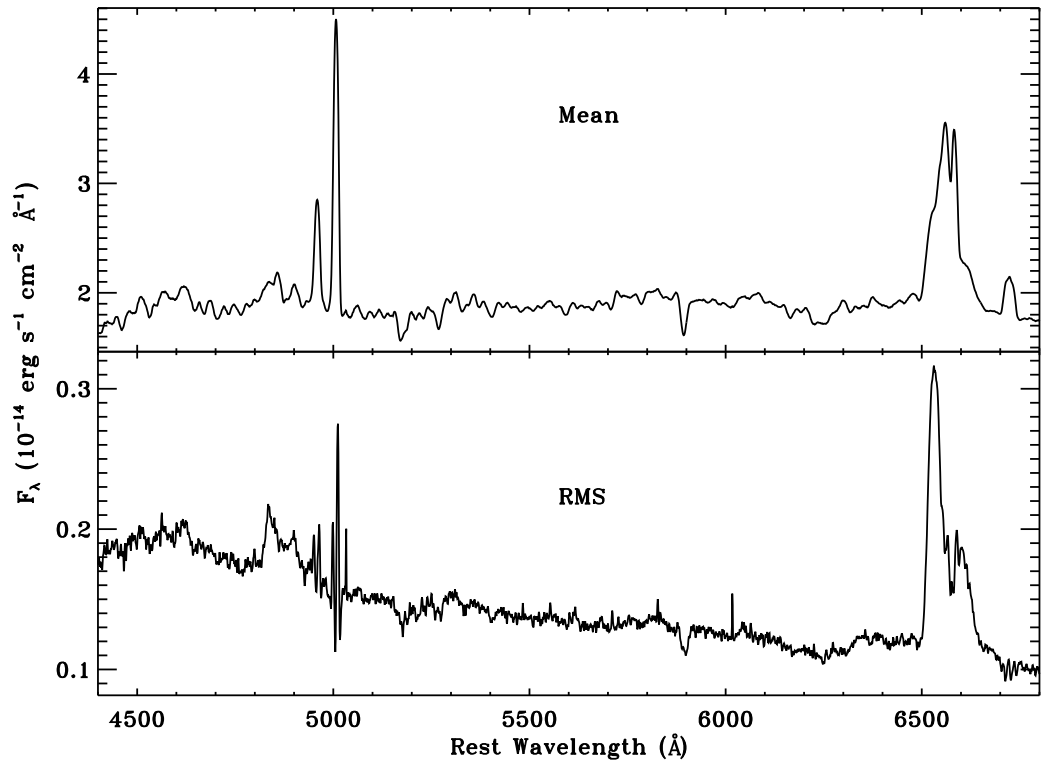}
 \end{center}
 \caption{The mean (top) and RMS (bottom) spectra.}
  \label{fig1}
\end{figure*}

\clearpage

\begin{figure*}
 \begin{center}
  \includegraphics[angle=0,scale=0.8]{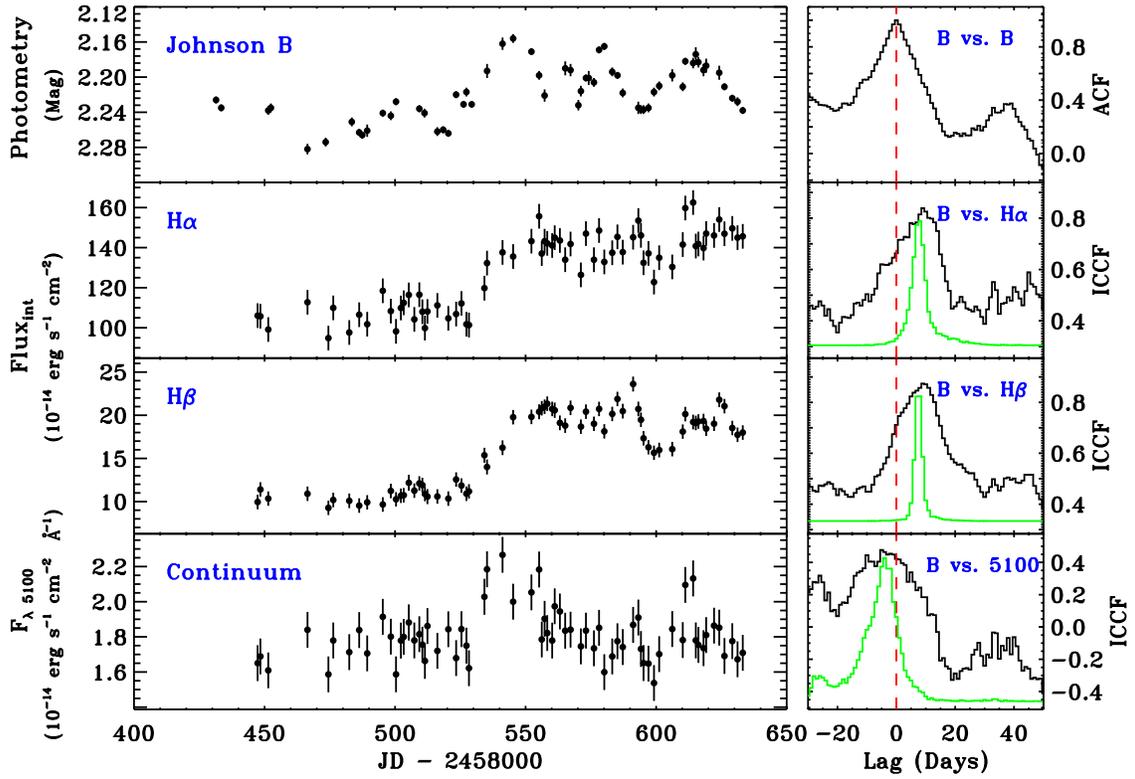}
 \end{center}
 \caption{LCs of photometry, H$\alpha$, H$\beta$, and 5100 \AA\ (the left panel). The LCs of H$\alpha$, H$\beta$, and 5100 \AA\ are the integrated LCs. The right penal is ACF, ICCF (black), and CCCD (green). Fluxes of H$\alpha$ and H$\beta$ are obtained by integrating the continuum-subtracted spectra containing the narrow lines. The continuum flux density is obtained by integrating the spectra.}
  \label{fig2}
\end{figure*}

\clearpage

\begin{figure*}
 \begin{center}
  \includegraphics[angle=0,scale=0.8]{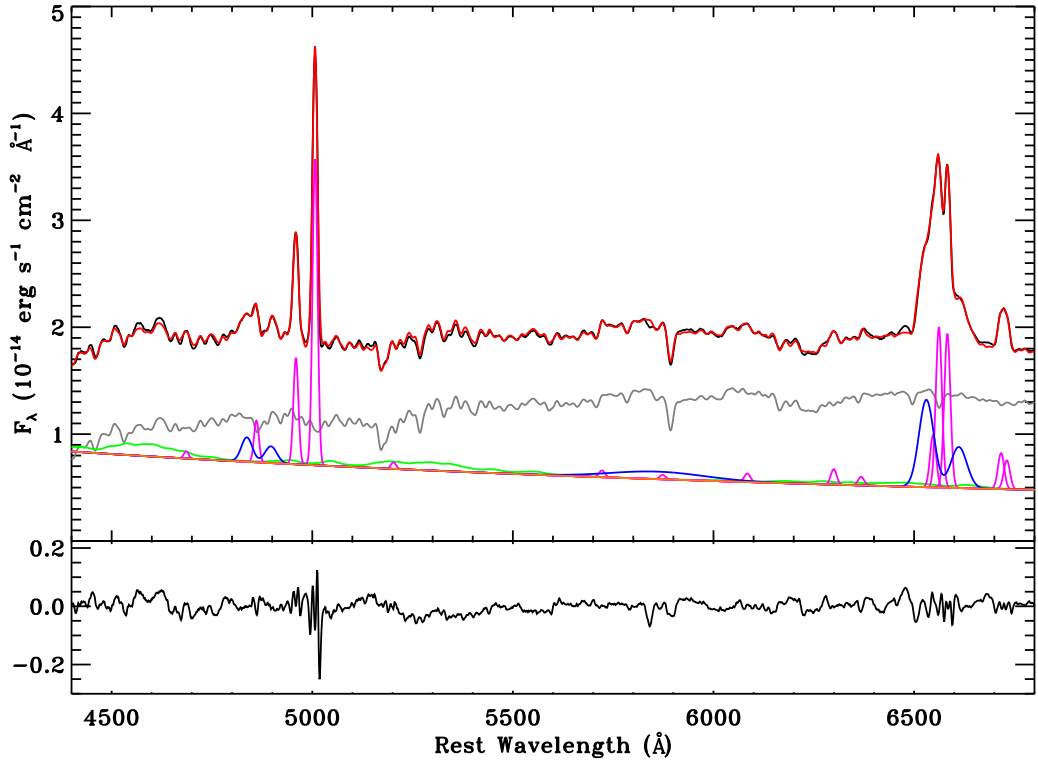}
 \end{center}
 \caption{The fitting results of the mean spectrum. The top panel shows the observed mean spectrum (black), the best-fitting component (red), host galaxy (grey), \feii\ emission (green), power-law continuum (orange), broad emission lines (blue), and narrow emission lines (magenta). The bottom shows the corresponding residuals.}
  \label{fig3}
\end{figure*}

\begin{figure*}
 \begin{center}
  \includegraphics[angle=0,scale=0.8]{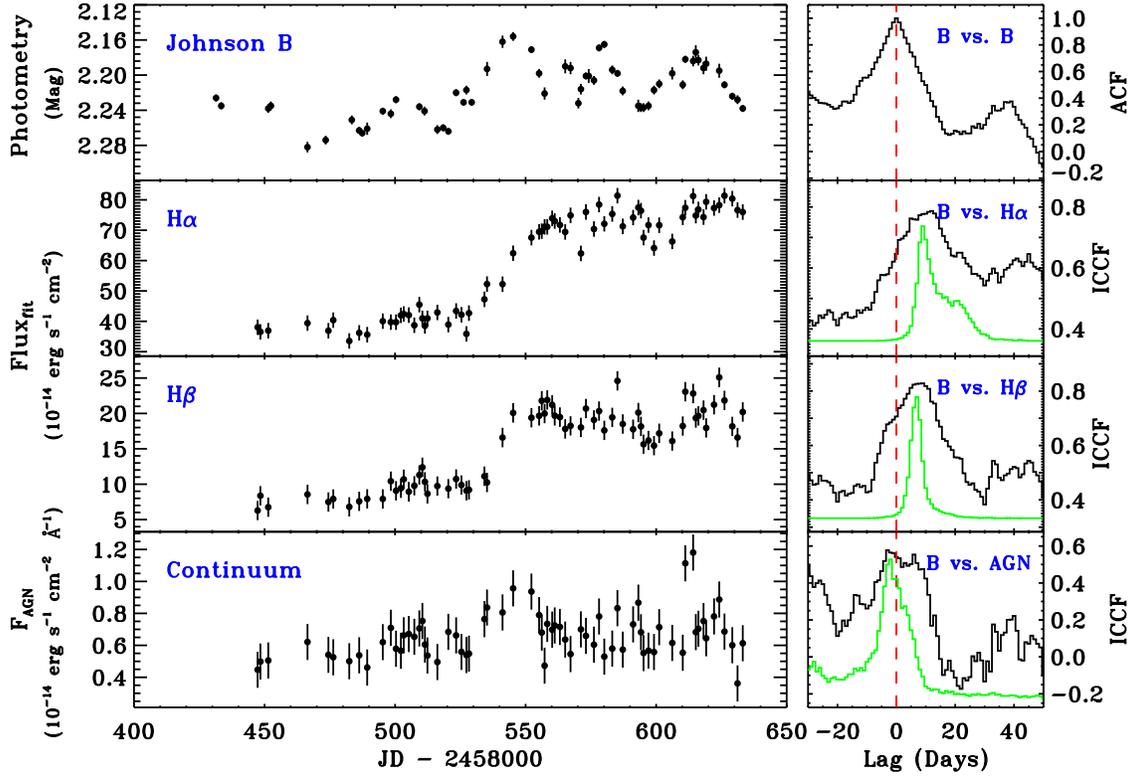}
 \end{center}
 \caption{LCs of photometry, H$\alpha$, H$\beta$, and continuum (the left panel). The fluxes of H$\alpha$ and H$\beta$ are measured from the best-fitting values of broad components, and the continuum is represented by the fitted power law around 5100 \AA. The LCs of H$\alpha$, H$\beta$, and 5100 \AA\ are the fitted LCs. The format is same as in Figure 2.}
  \label{fig4}
\end{figure*}

\begin{figure*}
 \begin{center}
  \includegraphics[angle=0,scale=1.0]{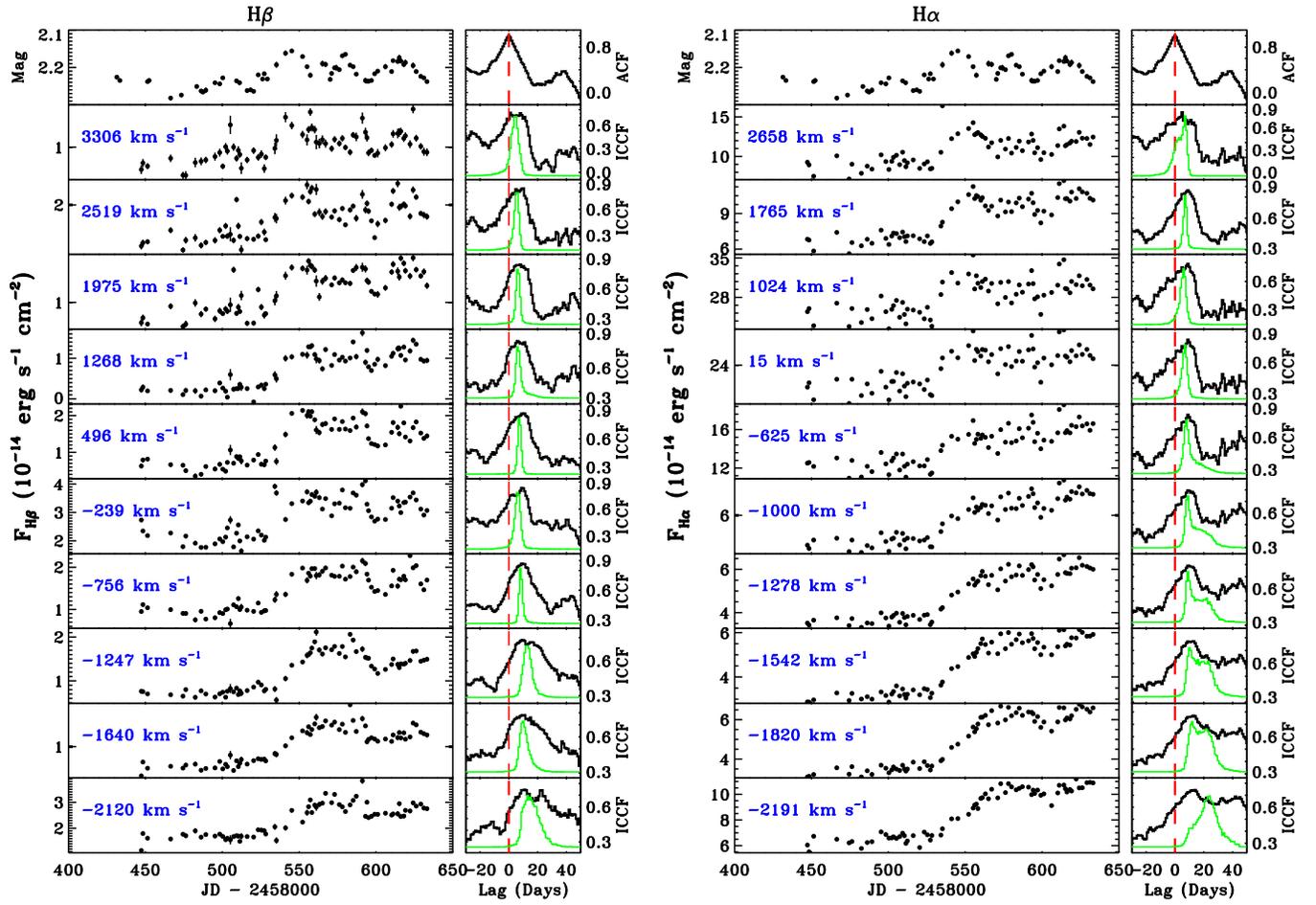}
 \end{center}
 \caption{The LCs in different velocity bins. The format is same as in Figure 2.}
  \label{fig5}
\end{figure*}

\begin{figure*}
 \begin{center}
  \includegraphics[angle=0,scale=0.8]{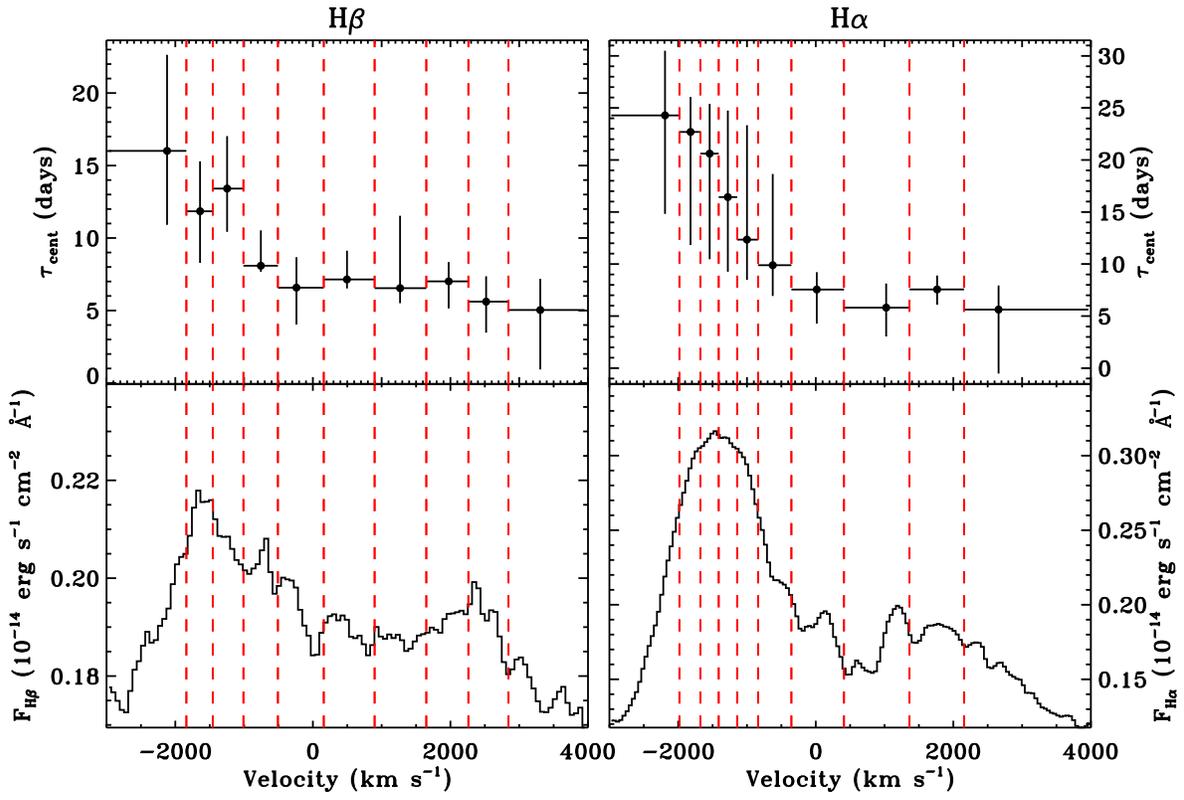}
 \end{center}
 \caption{The top panel shows the velocity-resolved delays of H$\alpha$ (right) and H$\beta$ (left). The bottom panel shows the RMS spectrum of of H$\alpha$ (right) and H$\beta$ (left).}
  \label{fig6}
\end{figure*}


\begin{thebibliography}{99}

\bibitem[Ai et al. (2020)]{Al20} Ai, Y. L., Dou, L. M., Yang, C. W,, et al. 2020, ApJL, 890, L29

\bibitem[Alexander (1997)]{Al97} Alexander, T. 1997, in Astronomical Time Series, ed. D. Maoz, A. Sternberg, \& E. M. Leibowitz (Dordrecht: Kluwer), 163

\bibitem[Andrillat (1968)]{An68} Andrillat, Y. 1968, AJ, 73, 862

\bibitem[Antonucci (1993)]{An93} Antonucci, R. 1993, ARA\&A, 31, 473

\bibitem[Antonucci \& Miller (1985)]{AM85} Antonucci, R. R. J., \& Miller, J. S. 1985, ApJ, 297, 621

\bibitem[Aretxaga et al. (1999)]{Ar99} Aretxaga, I., Joguet, B., Kunth, D., et al. 1999, ApJ, 519, 123

\bibitem[Barth et al. (2015)]{Ba15} Barth, A. J., Bennert, V. N., Canalizo, G., et al. 2015, ApJS, 217, 26

\bibitem[\protect\citeauthoryear{Bentz et al.}{2013}]{Be13} Bentz, M. C., Denney, K. D., Grier, C. J., et al. 2013, ApJ, 767, 149

\bibitem[\protect\citeauthoryear{Bentz et al.}{2009a}]{Be09a} Bentz, M. C., Peterson, B. M., Netzer, H., Pogge, R. W., \& Vestergaard, M. 2009a, ApJ, 697, 160

\bibitem[\protect\citeauthoryear{Bentz et al.}{2009b}]{Be09b} Bentz, M. C., Walsh, J. L., Barth, A. J., et al. 2009b, ApJ, 705, 199

\bibitem[Bentz et al. (2010)]{Be10} Bentz, M. C., Walsh, J. L., Barth, A. J., et al. 2010, ApJ, 716, 993

\bibitem[Bianchi et al. (2005)]{Bi05} Bianchi, S., Guainazzi, M., Matt, G., et al. 2005, A\&A, 442, 185

\bibitem[Blandford \& McKee (1982)]{BM82} Blandford, R. D., \& McKee, C. F. 1982, ApJ, 255, 419

\bibitem[Bruzual \& Charlot (2003)]{BC03} Bruzual, G., \& Charlot, S. 2003, MNRAS, 344, 1000

\bibitem[Denney et al. (2014)]{De14} Denney, K. D., De Rosa, G., Croxall, K., et al. 2014, ApJ, 796, 134


\bibitem[Denney et al. (2010)]{De10} Denney, K. D., Peterson, B. M., Pogge, R. W., et al. 2010, ApJ, 721, 715

\bibitem[De Rosa et al. (2018)]{De18} De Rosa, G., Fausnaugh, M. M., Grier, C. J., et al. 2018, ApJ, 866, 133

\bibitem[Devereux (2016)]{De16} Devereux, N. 2016, ApJ, 822, 69


\bibitem[Dong et al. (2008)]{Do08} Dong, X.-B., Wang, T.-G., Wang, J.-G., et al. 2008, MNRAS, 383, 581

\bibitem[Du et al. (2018)]{Du18a} Du, P., Brotherton, M. S., Wang, K., et al. 2018, ApJ, 869, 142

\bibitem[Du et al. (2014)]{Du14} Du, P., Hu, C., Lu, K., et al. 2014, ApJ, 782, 45

\bibitem[Dunn et al. (2018)]{Du18b} Dunn, J. P., Parvaresh, R., Kraemer, S. B., \& Crenshaw, D. M. 2018, ApJ, 854, 166

\bibitem[Eracleous \& Halpern (2001)]{EH01} Eracleous, M., \& Halpern, J. P. 2001, ApJ, 554, 240

\bibitem[\protect\citeauthoryear{Eracleous et al.}{1995}]{Er95} Eracleous, M., Livio, M, Halpern J. P., \& Storchi-Bergmann, T. 1995, ApJ, 438, 610

\bibitem[Esin et al. (1997)]{Es97} Esin, A. A., McClintock, J. E., Narayan, R. 1997, ApJ, 489, 865

\bibitem[Fausnaugh et al. (2017)]{Fa17} Fausnaugh, M. M., Grier, C. J., Bentz, M. C., et al. 2017, ApJ, 840, 97

\bibitem[Feng et al. (2017)]{Fe17} Feng, Hai-Cheng, Liu, H. T., Fan, X. L., et al. 2017, ApJ, 849, 161

\bibitem[Feng et al. (2020)]{Fe20} Feng, Hai-Cheng, Liu, H. T., Bai, J. M., et al. 2020, ApJ, 888, 30

\bibitem[Fitzpatrick (1999)]{Fi99} Fitzpatrick, E. L. 1999, PASP, 111, 63

\bibitem[Gezari et al. (2017)]{Ge17} Gezari, S., Hung, T., Cenko, S. B., et al. 2017, ApJ, 835, 144

\bibitem[Guo et al. (2020)]{Gu20} Guo, H., Shen, Y., He, Z., et al. 2020, ApJ, 888, 58

\bibitem[Hon et al. (2020)]{Ho20} Hon, W.-J., Webster, R., \& Wolf, C. 2020, MNRAS, 497, 192

\bibitem[Horne et al. (2004)]{Ho04} Horne, K., Peterson, B. M., Collier, S. J., \& Netzer, H. 2004, PASP, 116, 465

\bibitem[Hu et al. (2020)]{Hu20} Hu, C., Li, Y.-R., Du, P., et al. 2020, ApJ, 890, 71

\bibitem[Hu et al. (2016)]{Hu16} Hu, C., Wang, J.-M., Ho, L. C., et al. 2016, ApJ, 832, 197

\bibitem[Igarashi et al. (2020)]{Ig20} Igarashi, T., Kato, Y., Takahashi, H. R., et al. 2020, ApJ, 902, 103


\bibitem[Ili\'{c} et al. (2020)]{Il20} Ili\'{c}, D., Oknyansky, V., Popovi\'{c}, L. \v{C}., et al. 2020, A\&A, 638, A13

\bibitem[\protect\citeauthoryear{Kaspi et al.}{2000}]{Ka00} Kaspi, S., Smith, P. S., Netzer, H., et al. 2000, ApJ, 533, 631


\bibitem[Keel (1996)]{Ke96} Keel, W. C. 1996, AJ, 111, 696

\bibitem[Khachikian \& Weedman (1974)]{KW74} Khachikian, E. Y., \& Weedman, D. W. 1974, ApJ, 192, 581

\bibitem[Kim et al. (2018)]{Ki18} Kim, D. C., Yoon, I., Evans, A. S. 2018, ApJ, 861, 51

\bibitem[Kokubo \& Minezaki (2020)]{KM20} Kokubo, M., \& Minezaki, T. 2020, MNRAS, 491, 4615

\bibitem[Kollatschny (2020)]{Ko20} Kollatschny, W., Grupe, D., Parker, M. L., et al. 2020, A\&A, 638, A91

\bibitem[Korista \& Goad (2004)]{Ko04} Korista, K. T., \& Goad, M. R. 2004, ApJ, 606, 749

\bibitem[\protect\citeauthoryear{Kova\v{c}evi\'{c} et al.}{2018}]{Ko18} Kova\v{c}evi\'{c}, A. B., P\'{e}rez-Hern\'{a}ndez, E., Popovi\'{c}, L. \v{C}., et al. 2018, MNRAS, 475, 2051

\bibitem[Kraemer et al. (2002)]{Kr02} Kraemer, S. B., Crenshaw, D. M., \& George, I. M., et al. 2002, ApJ, 577, 98


\bibitem[LaMassa et al. (2015)]{La15} LaMassa, S. M., Cales, S., Moran, E. C., et al. 2015, ApJ, 800, 144

\bibitem[LaMassa et al. (2017)]{La17} LaMassa, S. M., Yaqoob, T., \& Kilgard, R., et al. 2017, ApJ, 840, 11


\bibitem[Liu et al. (2019)]{Li19} Liu, H. T., Feng H.-C., Xin, Y. X., et al. 2019, ApJ, 880, 155

\bibitem[Liu et al. (2017)]{Li17} Liu, H. T., Feng, H.-C., \& Bai, J. M. 2017, MNRAS, 466, 3323

\bibitem[Lu et al. (2016)]{Lu16} Lu, K. X., Du, P., Hu, C., et al. 2016, ApJ, 827, 118


\bibitem[MacLeod et al. (2019)]{Ma19} MacLeod, C. L., Green, P. J., Anderson, S. F., et al. 2019, ApJ, 874, 8

\bibitem[MacLeod et al. (2016)]{Ma16} MacLeod, C. L., Ross, N. P., Lawrence, A., et al. 2016, MNRAS, 457, 389

\bibitem[Mandal et al. (2020)]{Ma20} Mandal, A. K., Rakshit, S., Stalin, C. S., et al. 2020, MNRAS, accepted (arXiv: 2012.04906)

\bibitem[Matt et al. (2003)]{Ma03} Matt, G., Guainazzi, M., \& Maiolino, R. 2003, MNRAS, 342, 422

\bibitem[Merloni et al. (2015)]{Me15} Merloni, A., Dwelly, T., Salvato, M., et al. 2015, MNRAS, 452, 69

\bibitem[Miller \& Goodrich (1990)]{MG90} Miller, J. S., \& Goodrich R. W. 1990, ApJ, 335, 456



\bibitem[Netzer (1975)]{Ne75} Netzer, H. 1975, MNRAS, 171, 395

\bibitem[Noda et al. (2016)]{No16} Noda, H., Minezaki, T., Watanabe, M., et al. 2016, ApJ, 828, 78

\bibitem[\protect\citeauthoryear{Oknyansky et al.}{2019}]{Ok19} Oknyansky,V. L., Winkler, H., Tsygankov, S. S., et al. 2019, MNRAS, 483, 558

\bibitem[Onken et al. (2004)]{On04} Onken, C. A., Ferrarese, L., Merritt, D., et al. 2004, ApJ, 615, 645

\bibitem[Osterbrock (1981)]{Os81} Osterbrock, D. E. 1981, ApJ, 249, 462

\bibitem[Peterson (1993)]{Pe93} Peterson, B. M. 1993, PASP, 105, 247

\bibitem[\protect\citeauthoryear{Peterson et al.}{2002}]{Pe02} Peterson, B. M., Berlind, P., Bertram, R., et al. 2002, ApJ, 581, 197

\bibitem[Peterson et al. (2004)]{Pe04} Peterson, B. M., Ferrarese, L., Gilbert, K. M., et al. 2004, ApJ, 613, 682

\bibitem[Peterson et al. (1998)]{Pe98} Peterson, B. M., Wanders, I., Bertram, R., et al. 1998, ApJ, 501, 82

\bibitem[\protect\citeauthoryear{Piotrovich et al.}{2015}]{Pi15} Piotrovich, M. Y., Gnedin, Y. N., Silant'ev, N.
A., Natsvlishvili, T. M., \& Buliga, S. D. 2015, MNRAS, 454, 1157

\bibitem[Popovi\'{c}, et al. (2002)]{Po02} Popovi\'{c}, L. \v{C}., Mediavilla, E. G., Kubi\v{c}ela, A., \& Jovanovi\'{c}, P. 2002, A\&A, 390, 473


\bibitem[Pozo Nu\~{n}ez et al. (2015)]{Po15} Pozo Nu\~{n}ez, F., Ramolla, M., Westhues, C., et al. 2015, A\&A, 576, A73

\bibitem[Rakshit (2020)]{Ra20} Rakshit, S. 2020, A\&A, 642, A59

\bibitem[Rees et al. (1989)]{Re89} Rees, M. J., Netzer, H., \& Ferland, G. J. 1989, ApJ, 347, 640

\bibitem[Ricci et al. (2020)]{Ri20} Ricci, C., Kara, E., Loewenstein, M., et al. 2020, ApJL, 898, L1

\bibitem[Rumbaugh et al. (2018)]{Ru18} Rumbaugh, N., Shen, Y., Morganson, E., et al. 2018, ApJ, 854, 160



\bibitem[Schlegel et al. (1998)]{Sc98} Schlegel, D. J., Finkbeiner, D. P., \& Davis, M. 1998, ApJ, 500, 525

\bibitem[Seyfert (1943)]{Se43} Seyfert, C. K. 1943, ApJ, 97, 28

\bibitem[Shapovalova et al. (2019)]{Sh19} Shapovalova, A. I., Popovi\'{c}, L. \v{C}., Afanasiev, V. L., et al.  2019, MNRAS, 485, 4790

\bibitem[Shappee et al. (2014)]{Sh14} Shappee, B. J., Prieto, J. L., Grupe, D., et al. 2014, ApJ, 788, 48

\bibitem[Sniegowska et al. (2020)]{Sn20} Sniegowska, M., Czerny, B., Bon, E., et al. 2020, A\&A, 641, A167

\bibitem[Stern et al. (2018)]{St18a} Stern, D., McKernan, B., Graham, M. J., et al. 2018, ApJ, 864, 27


\bibitem[Storchi-Bergmann et al. (1993)]{St93} Storchi-Bergmann, T., Baldwin, J. A., \& Wilson, A. S. 1993, ApJ, 410, L11

\bibitem[Storchi-Bergmann et al. (2017)]{St17} Storchi-Bergmann, T., Schimoia, J. S., Peterson, B. M., et al. 2017, ApJ, 835, 236


\bibitem[Sturm et al. (2018)]{St18b} Sturm, E., Dexter, J., Pfuhl, O., et al. 2018, Natur, 563, 657

\bibitem[Trakhtenbrot et al. (2019)]{Tr19} Trakhtenbrot, B., Arcavi, I., MacLeod, C. L., et al. 2019, ApJ, 883, 94


\bibitem[Tran et al. (1992)]{Tr92} Tran, H. D., Miller, J. S., \& Kay, L. E. 1992, ApJ, 397, 452

\bibitem[Tran (2003)]{Tr03} Tran, H. D. 2003, ApJ, 583, 632

\bibitem[\protect\citeauthoryear{Tremaine et al.}{2002}]{Tr02} Tremaine, S., Gebhardt, K., Bender, R., et al. 2002, ApJ, 574, 740

\bibitem[van Groningen \& Wanders (1992)]{vW92} van Groningen, E., \& Wanders, I. 1992, PASP, 104, 700

\bibitem[Wanders et al. (1993)]{Wa93} Wanders, I., van Groningen, E., Alloin, D., et al. 1993, A\&A, 269, 39

\bibitem[Wang et al. (2019)]{Wa19} Wang, C. J., Bai, J. M., Fan, Y. F., et al. 2019, RAA, 19, 149

\bibitem[Wang et al. (2018)]{Wa18} Wang, J., Xu, D. W., Wei, J. Y. 2018, ApJ, 858, 49

\bibitem[White \& Peterson (1994)]{WP94} White, R. J., \& Peterson, B. M. 1994, PASP, 106, 879

\bibitem[Whittle (1992)]{Wh92} Whittle, M. 1992, ApJS, 79, 49

\bibitem[\protect\citeauthoryear{Woo et al.}{2013}]{Wo13} Woo, J. H., Schulze, A., Park, D., et al. 2013, ApJ, 772, 49

\bibitem[Woo et al. (2010)]{Wo10} Woo, J. H., Treu, T., Barth, A. J., et al. 2010, ApJ, 716, 269

\bibitem[\protect\citeauthoryear{Woo et al.}{2015}]{Wo15} Woo, J. H., Yoon, Y., Park, S., Park, D., \& Kim, S.
C. 2015, ApJ, 801, 38

\bibitem[Wu et al. (2011)]{Wu11} Wu, Y. Z., Zhang, E. P., Liang, Y. C., et al. 2011, ApJ, 730, 121

\bibitem[Xin et al. (2020)]{Xi20} Xin, Y. X., Bai, J. M., Lun, B. L., et al. 2020, RAA, 20, 149

\bibitem[Yang et al. (2018)]{Ya18} Yang, Q., Wu, X.-B., Fan, X., et al. 2018, ApJ, 862, 109

\bibitem[Yuan \& Narayan (2014)]{YN14} Yuan, F., \& Narayan, R. 2014, ARA\&A, 52, 529

\bibitem[Zetzl et al. (2018)]{Ze18} Zetzl, M., Kollatschny, W., Ochmann, M. W., et al. 2018, A\&A, 618, 83




\end{thebibliography}
\end{document}